# I-DREM MRAC with Time-Varying Adaptation Rate & No A Priori Knowledge of Control Input Matrix Sign to Relax PE Condition*

Glushchenko A.I., *Member, IEEE*, Petrov V.A., and Lastochkin K.A.

*Abstract*—The known dynamic regressor extension and mixing method (DREM) is combined with the proposed filter of a new type, which uses the integration operation with forgetting, and the recursive least-squares method to develop the new I-DREM model reference adaptive control (MRAC) system. It does not require a priori knowledge of the sign or the elements values of the control input matrix of the plant. It also provides the exponential convergence of the adaptation process (with the automatically adjustable adaptation rate) without the regressor persistent excitation. Such control system allows to solve three actual problems of the adaptive control: 1) to provide the exponential convergence of the controller parameter error under the condition of the regressor initial excitation, 2) to make such convergence monotonic, 3) to calculate the adaptation rate online according to the current regressor value. Some numerical experiments are conducted to demonstrate the effectiveness of the proposed method.

## I. Introduction

Methods of the model reference adaptive control (MRAC) have proved to be effective for the synthesis of control systems in the cases when the influence of parametric uncertainties on control quality is strong [1]. In contrast to the methods of linear, robust, and optimal control theories, the MRAC methods provide the required control quality, rather than the artificially lowered one, in the case of the sufficient control margin. This is because the aim of the MRAC system is to make the plant output coincide with the reference model one. The main assumption is that there is a set of ideal parameters of the chosen control law, which ensures that this aim is attainable [1, 2]. Thus the task of synthesis of the MRAC control system comes down to the development of the adaptation loop to adjust the control law parameters. There are methods of direct and indirect MRAC with the standard loops of adaptation of $\Gamma e^T PBx$ type, which have been known for more than half a century. Further, we will consider in detail the main problems, which one faces when the above-mentioned adaptation loops are used.

The first problem of the standard adaptation loops is the asymptotic convergence of the tracking error, but not the parameter error of the control law. The fact is that the convergence of the parameters to their ideal values is a very advantageous property for MRAC [3-5] because it automatically leads to the exponential convergence of the tracking error too [5], as well as the robustness of the obtained control loop. When the parameter error convergence is not provided, such a loop requires the application of the *e*- and σ- modifications or the projection operator [2, 6]. For the standard adaptation loop, the parameters convergence of the control law is guaranteed only if the condition of the regressor persistent excitation (PE) is satisfied [1, 2]. In [7] it has been proved that the PE condition is met for MRAC if the number of spectral lines in a given reference signal coincides with the number of the unknown controller parameters. It is obvious that, for many practical applications, the fulfillment of this condition leads to increased energy consumption and wear and tear of actuators. Therefore, a number of solutions [8-10] have been proposed to relax the PE condition. First of all, this is the Switched MRAC [8] using a PI-like adaptation loop [9], and, secondly, the Concurrent Learning [10], which uses an array of retrospective data stored in a DataStack along with the current measurements. Both methods allow one to relax the PE condition to the initial excitation one (IE), which is satisfied for almost any type of reference signal. The disadvantages of the first approach are the non-formalized stability analysis at switching moments and the application of open-loop integration. As for the second method, these are questions on how to form the DataStack and the need to know the values of the elements of the unmeasurable vector of the tracking error first derivative [11].

The second problem of the standard loops is the need to know the sign/value of the plant control input matrix [2]. The essence of this problem is the possible singularity of the solutions when the parameter of the control law feedforward part crosses zero as a result of its adjustment. This problem occurs because the used parameterization requires this parameter to be inverted [2]. To solve the singularity problem, a nonlinear operator has been proposed in [2] that allows keeping the feedforward controller parameter away from a given range around zero value. However, in practice, for the standard adaptation loops, this method often leads to chattering between the boundaries of such a range. In the recent paper [12], a solution of the chattering problem has been proposed. It is obtained by the application of an adaptation loop based on the DREM procedure [13]. It allows to provide monotonic convergence of the adjustable parameters to guarantee only one switching of the nonlinear operator [12] in the course of the adaptation.

The third problem with the standard adaptation loops is the constant experimentally chosen value of the adaptation rate $\Gamma$ [1, 2, 14]. The well-known papers [2, 6] state that the high values of $\Gamma$ could result in the excitation of the plant high-frequency unmodelled dynamics, and also in the instability in the case of even limited disturbances. Therefore, modifications of MRAC, such as the frequency limited

* Research was financially supported by Russian Foundation for Basic Research (Grant 18-47-310003-r_a).

A. I. Glushchenko is with Stary Oskol technological institute (branch) NUST "MISIS", Stary Oskol, Russia (phone: +79102266946; e-mail: a.glushchenko@sf-misis.ru).

V. A. Petrov is with Stary Oskol technological institute (branch) NUST "MISIS", Stary Oskol, Russia (e-mail: petrov.va@misis.ru).

K. A. Lastochkin is with Stary Oskol technological institute (branch) NUST "MISIS", Stary Oskol, Russia (e-mail: lastconst@yandex.ru).

adaptive control [15], the L₁ adaptive control [16], the MRE and DRE schemes [11] have been proposed to allow one to use the high values of Γ. However, these methods do not solve the problem of manual selection of Γ. In practice, experiments to choose it often cannot be conducted, particularly because of the fact that its optimal value depends on the type and value of the reference signal and the current excitation of the regressor. To solve this problem, methods have been proposed in [17, 18], which allow scaling the once chosen rate Γ depending on the current values of the regressor and the reference. Also in [19], instead of the standard adaptation loop, it is proposed to develop the one on the basis of the recursive least-squares (RLS) method, which provides a law to adjust Γ. However, to be implemented, the method in [19] requires the first derivative of the tracking error to be measured. Also, RLS in [19] is used without the exponential forgetting factor. This fact causes some well-known problems [1, 2].

Thus, all three problems of the standard adaptation loops have a long history and interesting solutions, even a small part of which has not been mentioned in this section. However, as a rule, solving one of these problems, all the rest are omitted. For example, the Concurrent Learning [10], which relaxes the PE condition, requires the control input matrix to be known. On the contrary, the methods [12], which do not have such a restriction or adjust the adaptation rate [17, 18], require the PE condition to be met to provide the exponential convergence.

Therefore, *the novelty and contribution* of this research is the development of a loop of adaptation of the control law parameters, which considers all three described problems of the standard loops at once. For this purpose, it is suggested to filter the equation of error between the outputs of the plant and the reference model in a special way. As a result, the parameter uncertainty will be explicitly defined. Then the DREM procedure [13] will be applied to exclude the assumption about the knowledge of the control input matrix, like in [12]. Finally, using the MRE procedure [11], the scalarly parameterized parameter uncertainty will be passed through the filter proposed in [20]. This will allow us to apply the recursive least squares method with the exponential forgetting factor to adjust the controller parameters and will provide exponential parameter convergence when the IE condition is met.

## II. I-DREM MRAC

### A. Parameterization

Let a problem of control of one class of the linear time-invariant (LTI) plants be considered:

$$\dot{x} = Ax + Bu, \quad (1)$$

where $x \in R^n$ is a plant state vector, $u \in R^m$ is a control action, $A \in R^{n \times n}$ is a system state matrix, and $B \in R^{n \times m}$ is a control input matrix of full column rank. The values of elements of $A$ and $B$ are unknown, but $(A, B)$ is a controllable pair, so as $m \leq n$. The state vector $x$ is directly measurable.

The reference model, which defines the required control quality for the closed-loop system with the control action $u$ and the plant (1), is defined as:

$$\dot{x}_{ref} = A_{ref} x_{ref} + B_{ref} r, \quad (2)$$

where $x_{ref} \in R^n$ is a state vector of the reference model, $r \in R^m$ is a setpoint signal, $B_{ref} \in R^{n \times m}$ is a reference input matrix of full column rank. The reference model state matrix $A_{ref} \in R^{n \times n}$ is a Hurwitz one.

The control law for the plant (1) is defined as:

$$u = \hat{k}_r \hat{k}_x x + \hat{k}_r r. \quad (3)$$

$\hat{k}_x \in R^{m \times n}$, $\hat{k}_r \in R^{m \times m}$ are its adjustable parameters. Let (3) be substituted into (1) to obtain the closed-loop equation:

$$\dot{x} = \left( A + B \hat{k}_r \hat{k}_x \right) x + B \hat{k}_r r. \quad (4)$$

**Assumption 1.** *There exist such ideal values of the parameters $k_x \in R^{m \times n}$ and $k_r \in R^{m \times m}$ of the control law that the following equalities hold:*

$$A + B k_r k_x = A_{ref}; \quad B k_r = B_{ref}. \quad (5)$$

If $A$ and $B$ have the same structures as $A_{ref}$ and $B_{ref}$ respectively, then Assumption 1 is met [1, 2].

Then the error equation is formed as the difference between $x$ and $x_{ref}$:

$$\begin{aligned}
\dot{e}_{ref} &= A_{ref} e_{ref} + B \left[ \hat{k}_r \hat{k}_x x + \hat{k}_r r \right] - \left( A_{ref} - A \right) x - B_{ref} r = \\
&= A_{ref} e_{ref} + B_{ref} \left( k_r^{-1} \left[ \hat{k}_r \hat{k}_x x + \hat{k}_r r \right] - k_x x - r \pm \hat{k}_x x \right) = \\
&= A_{ref} e_{ref} + B_{ref} \left[ \tilde{k}_x x + \left( k_r^{-1} \hat{k}_r - I \right) \left( \hat{k}_x x + r \right) \right] = \\
&= A_{ref} e_{ref} + B_{ref} \left[ \tilde{k}_x x - \tilde{k}_r^{-1} \hat{k}_r \left( \hat{k}_x x + r \right) \right].
\end{aligned} \quad (6)$$

Here $\tilde{k}_x = \hat{k}_x - k_x$, $\tilde{k}_r^{-1} = \hat{k}_r^{-1} - k_r^{-1}$. Let the following matrix notation be introduced into (6):

$$\varphi = \left[ x^T \quad -\left( \hat{k}_x x + r \right)^T \hat{k}_r \right]^T; \quad \tilde{\theta}^T = \left[ \tilde{k}_x \quad \tilde{k}_r^{-1} \right] = \hat{\theta}^T - \theta^T, (7)$$

where $\varphi \in R^{n+m}$, $\tilde{\theta}^T \in R^{m \times (n+m)}$. Taking into consideration (7), the equation (6) can be rewritten as:

$$\dot{e}_{ref} = A_{ref} e_{ref} + B_{ref} \tilde{\theta}^T \varphi. \quad (8)$$

The parameterization (8) has been proposed in [21] to obtain the adaptation loop, which does not require the *a priori* knowledge of the control input matrix $B$ of the plant. However, the adaptive laws derived in [21] require the sign of $B$ to be known, as we need to invert $\hat{k}_r$ to obtain $\hat{k}_r^{-1}$ to implement the control law (3). In the case of the disturbances or the incorrect choice of the initial value of sign of $\hat{k}_r^{-1}$, this will result in $\hat{k}_r$ zero (the singularity point for $\hat{k}_r^{-1}$) crossing in the course of its adjustment. In this research, using the parameterization (8), the adaptive law will be derived, which does not require the *a priori* knowledge of values of elements or the sign of $B$.

### B. Filtration to estimate error vector first derivative

Using the filtration procedure, let the parameter uncertainty function be explicitly written in (8). For this

purpose, let the notion of a compensatory control $u_c$ be introduced into (8):

$$\dot{e}_{ref} = A_{ref}e_{ref} + B_{ref}\left(u_c - \theta^T\varphi\right). \quad (9)$$

All the dynamic variables of (9) are to be filtrated using the aperiodic links:

$$\dot{\mu}_f = -l\mu_f + \dot{e}_{ref}, \ \mu_f(0) = 0; \quad (10)$$
$$\dot{e}_f = -le_f + e_{ref}, \ e_f(0) = 0;$$
$$\dot{u}_{cf} = -lu_{cf} + u_c, \ u_{cf}(0) = 0; \quad (11)$$
$$\dot{\varphi}_f = -l\varphi_f + \varphi, \ \varphi_f(0) = 0,$$

where $l > 0$ is a filter time constant.

The values of $e_f$, $\varphi_f$, and $u_{cf}$ are calculated as solutions of the respective differential equations (10), (11). The value $\mu_f$ is calculated using the following lemma.

**Lemma 1.** *The filtered value $\mu_f$ of the first derivative of the tracking error can be calculated as follows, using the measurable signals:*

$$\mu_f = e^{-lt}\mu_f(0) + e_{ref}(t) - e^{-lt}e_{ref}(0) - le_f(t) + le^{-lt}e_f(0). \quad (12)$$

*The proof of Lemma 1 is provided in Part A of Appendix.*

Considering the filtration (10), (11), the error equation (9) is rewritten as:

$$\mu_f = A_{ref}e_f + B_{ref}\left(u_{cf} - \theta^T\varphi_f\right), \ \forall t \geq 0. \quad (13)$$

Let the required behavior of (13) be introduced as:

$$\mu_{fd} = A_{ref}e_f + B_{ref}u_{cf}. \quad (14)$$

The parameter disturbance for the error equation (8) can be calculated as the difference between $\mu_{fd}$ and $\mu_f$:

$$\mu_{fd} - \mu_f = B_{ref}\theta^T\varphi_f. \quad (15)$$

Let the equation (15) be rewritten as the classical linear regression equation (LRE):

$$y = B_{ref}^\dagger\left(\mu_{fd} - \mu_f\right) = \theta^T\varphi_f, \quad (16)$$

where $y \in R^m$ is the parameter uncertainty function, $\varphi_f \in R^{n+m}$ is the regressor vector.

*C. DREM*

Using the DREM procedure [13], the matrix regressor $\varphi_f$ (an individual regressor for each row of $\theta$) is transformed into the scalar one, which is common for all rows of $\theta$. Let the $n+m-1$ stable operators be introduced:

$$(.)_{f_i(t)} := \left[H_i(.)\right](t); \ H_i(p) = \frac{\alpha_i}{p + \beta_i}; \quad (17)$$
$$i \in \{1, 2, \cdots, n + m - 1\},$$

where $\alpha_i > 0$ and $\beta_i > 0$ are filters constants.

Having passed the function $y$ and the regressor $\varphi_f$ through the operators (17), the extended LRE is formed:

$$Y_f(t) = \Phi_f(t)\theta,$$
$$Y_f(t) = \left[y(t) \ y_{f_1}(t) \ \cdots \ y_{f_{n+m-1}}(t)\right]^T; \quad (18)$$
$$\Phi_f(t) = \left[\varphi_f(t) \ \varphi_{ff_1}(t) \ \cdots \ \varphi_{ff_{n+m-1}}(t)\right]^T.$$

Then both left and right parts of the equation (18) are pre-multiplied by the adjugate matrix of the extended regressor $\Phi_f(t) - adj\{\Phi_f(t)\}$. As $adj\{\Phi_f(t)\}\Phi_f(t) = det\{\Phi_f(t)\}I$, we obtain:

$$Y(t) = adj\{\Phi_f(t)\}Y_f(t) = det\{\Phi_f(t)\}\theta = \omega(t)\theta, \quad (19)$$

where $Y \in R^{(n+m)\times m}$, $\omega \in R$ is the scalar regressor.

*D. I-DREM*

According to the DREM procedure [11, 13], the scalar gradient laws of $\theta$ elements estimation can be obtained using the regression (19), but $\omega \notin L_2$ is required for the monotonic asymptotic convergence. This is true if $\omega \in PE$ or for a small class of regressors $\omega \notin PE$ [13]. Therefore, to obtain an adaptation loop providing exponential convergence, firstly, we will apply the MRE procedure [11] and the filter proposed in [20] to the equations of regression (19). This will guarantee the monotonic exponential convergence of estimates of the $\theta$ elements in the case of $\omega \in IE$.

**Definition 1:** *The regressor $\omega(t) \in L_\infty$ is initially exciting ($\omega \in IE$) during the interval $[t_0; T_0]$, if for $T_0 > 0 \ \exists\alpha > 0$ such that*:

$$\int_{t_0}^{t_0+T_0}\omega^2(\tau)d\tau \geq \alpha, \quad (20)$$

where $\alpha$ is the degree of excitation, $t_0 > 0$ is the time moment, when, after filtration (17), $\omega(t)$ becomes non-zero.

**Remark 1.** *Because of the filters (17) stability and the fact that all of them are different, if the initial regressor $\varphi_f \in IE$, then the scalar regressor $\omega \in IE$ [22].*

Following the results of [20], the MRE method is applied to the scalar regression (19) [11]. For this purpose, the equation (19) is multiplied by the regressor $\omega(t)$:

$$\omega(t)Y_i(t) = \omega^2(t)\theta_i. \quad (21)$$

Let the filter from [20] be considered:

$$u_f(t) = \int_{t_0}^{t}e^{-\sigma\cdot\tau}u(\tau)d\tau, \quad (22)$$

where $\sigma > 0$ is the memory factor, $u_f$ and $u$ are the filter output and input respectively. The filter (22) is applied to the regressor $\omega^2$ and the function $\omega Y$:

$$\Omega(t) = \int_{t_0}^{t}e^{-\sigma\tau}\omega^2(\tau)d\tau; \ \Upsilon(t) = \int_{t_0}^{t}e^{-\sigma\tau}\omega(\tau)Y(\tau)d\tau. \quad (23)$$

**Proposition 1.** *If $\omega \in IE$, then the scalar regressor $\Omega(t)$ is positive semi-definite bounded function ($\Omega(t) \in L_\infty$ and $\Omega(t) \geq 0 \ \forall t \geq t_0$), such that: 1) $\Omega(t)$ increases to a finite limit $\Omega_{max}$; 2) $\Omega(t) \notin L_2$ [20]. Taking into consideration (23), the regression (19) is rewritten as:*

$$\Upsilon(t) = \Omega(t)\theta, \quad (24)$$

where $\Upsilon \in R^{(n+m) \times m}$, $\Omega \in R$. Let two parts be defined in the obtained equation (24):

$$\Upsilon^{k_x}(t) = \Omega(t) k_x; \quad \Upsilon^{k_r^{-1}}(t) = \Omega(t) k_r^{-1}. \quad (25)$$

As the following equalities hold for all $Q \in R^{m \times m}$ and $q \in R$

$$adj\{q \cdot Q\} = q \cdot adj\{Q\}, \quad det\{q \cdot Q\} = q^m \cdot det\{Q\}, \quad (26)$$

then we can derive the following from (25):

$$\begin{aligned}
M_a &= \Omega(t) N_a, \quad M_d = \Omega^m(t) N_d, \\
M_a(t) &= adj\{\Upsilon^{k_r^{-1}}(t)\}, \quad N_a = adj\{k_r^{-1}\}, \\
M_d(t) &= det\{\Upsilon^{k_r^{-1}}(t)\}, \quad N_d = det\{k_r^{-1}\}.
\end{aligned} \quad (27)$$

Then, taking into consideration the properties of the regressor $\Omega$, the adaptation laws for the adjugate matrix $N_a$, the determinant $N_d$ and the matrix $k_x$ are derived by application of the MRE procedure and the recursive least-squares method:

$$\begin{cases}
\dot{\hat{k}}_x = \Gamma \Omega^m \left[ \Omega^\rho \Upsilon^{k_x} - \Omega^m \hat{k}_x \right], \\
\dot{\hat{N}}_d = \Gamma \Omega^m \left[ M_d - \Omega^m \hat{N}_d \right], \\
\dot{\hat{N}}_a = \Gamma \Omega^m \left[ \Omega^\rho M_a - \Omega^m \hat{N}_a \right], \quad \dot{\Gamma} = \lambda \Gamma - \Gamma \Omega^{2m} \Gamma.
\end{cases} \quad (28)$$

where $\Gamma > 0$ is the adaptation rate, $\rho = m-1$ and $\lambda$ is the forgetting factor.

Considering (28), the equations for $\tilde{k}_x$ and $\tilde{N}_a, \tilde{N}_d$ are:

$$\dot{\tilde{k}}_x = -\Gamma \Omega^{2m} \tilde{k}_x; \quad \dot{\tilde{N}}_a = -\Gamma \Omega^{2m} \tilde{N}_a; \quad \dot{\tilde{N}}_d = -\Gamma \Omega^{2m} \tilde{N}_d. \quad (29)$$

As $\Gamma > 0$, then each element of $\tilde{k}_x$ and $\tilde{N}_a, \tilde{N}_d$ decreases monotonously.

*E. Relaxation of assumption of a priori knowledge of B*

Using (28), $\hat{k}_r$ could be calculated analytically with the help of the following equality $Q^{-1} = adj\{Q\} \cdot det^{-1}\{Q\}$:

$$\hat{k}_r = \hat{N}_d^{-1} \hat{N}_a. \quad (30)$$

$\hat{N}_d$ is to be prevented from zero crossing not to face singularity of (30). To implement such a restriction, the following non-linear operator is introduced [12]:

$$\hat{N}_d^{-1} = \begin{cases} \hat{N}_d^{-1}, & \text{if } |\hat{N}_d| > \underline{N}_d, \\ -\underline{N}_d^{-1} \text{sign}(\hat{N}_d), & \text{otherwise,} \end{cases} \quad (31)$$

where $\underline{N}_d > 0$ is the parameter, which defines the lower bound of $\hat{N}_d$.

**Proposition 2.** As $\hat{N}_d$ is monotonous, then: 1) if $sign(\hat{N}_d(0)) = sign(N_d)$, then there will be no switching (31) during adjustment; 2) if $sign(\hat{N}_d(0)) \neq sign(N_d)$, then only one switching is possible [12]. So if (31) is applied, then $\hat{k}_r$ is "protected" from becoming zero. As a result, $\hat{k}_r^{-1}$ is non-singular and can be calculated to be used in (11).

**Remark 2.** *Contrary to the known methods, which use the operator (31), in this research no chattering of $\hat{N}_d$ is guaranteed thanks to the monotonicity property. As for the requirement to know the matrix B, it has been firstly relaxed in [12] using the above-mentioned property.*

The laws (28) are rewritten in the vectorized form:

$$\begin{aligned}
\dot{\hat{\theta}}_{vec} &= \Gamma \Omega^m \left[ \Upsilon_{vec} - \Omega^m \hat{\theta}_{vec} \right], \quad \dot{\Gamma} = \lambda \Gamma - \Gamma \Omega^{2m} \Gamma, \\
\Upsilon_{vec}(t) &= \left[ vec^T(\Omega^\rho \Upsilon^{k_x}) \quad vec^T(\Omega^\rho M_a) \quad M_d \right]^T, \\
\hat{\theta}_{vec} &= \left[ vec^T(\hat{k}_x) \quad vec^T(\hat{N}_a) \quad \hat{N}_d \right]^T,
\end{aligned} \quad (32)$$

where $vec(.)$ is the vectorization operation. The properties of the error equation (8) combined with the adaptation loop (32) are defined as a theorem.

**Theorem 1.** If $\omega \in IE$ and the adaptation loop is defined as (32), then $\xi = \begin{bmatrix} e_{ref}^T & \tilde{\theta}_{vec}^T \end{bmatrix}^T$ exponentially converges to zero with an arbitrarily high rate.

The proof of Theorem 1 is provided in Part B of Appendix.

III. EXPERIMENTS

To validate the theoretical results and demonstrate the properties of the developed I-DREM MRAC method, the simulation of a closed-loop system of adaptive control (8) has been conducted. The control law parameters (3) were adjusted according to (32). The plant (1), the reference model (2) and setpoint $r$ were chosen as:

$$A = \begin{bmatrix} 1 & 1 \\ 4 & 2 \end{bmatrix}; \quad B = \begin{bmatrix} 1 & 0 \\ 0 & 1 \end{bmatrix}; \\
A_{ref} = \begin{bmatrix} 0 & 1 \\ -8 & -4 \end{bmatrix}; \quad B_{ref} = \begin{bmatrix} 4 & 2 \\ 0 & 2 \end{bmatrix}; \quad r = \begin{bmatrix} 1 \\ e^{-0.01t} \end{bmatrix}. \quad (33)$$

The values of the initial conditions, the control law (3) and the adaptation loop (32) parameters were chosen as:

$$\hat{k}_x(0) = 0_{2 \times 2}; \quad \hat{N}_a(0) = I_{2 \times 2}; \quad \hat{N}_d(0) = -0.125; \\
\Gamma = 0.1I; \quad \lambda = 1000. \quad (34)$$

The values of the parameters of the filters (10), (11), (17), (22) and the non-linear operator (31) were set as $l=100$; $\alpha_1=\alpha_2=\alpha_3=1$; $\beta_1=1$; $\beta_2=2$; $\beta_3=3$; $\sigma=0.125$; $\underline{N}_d = 0.025$. Figure 1 shows the norm of errors $\tilde{N}_a$ and $\tilde{k}_x$. Figure 2 shows the ideal value of $N_d$ and the transient of its estimation.

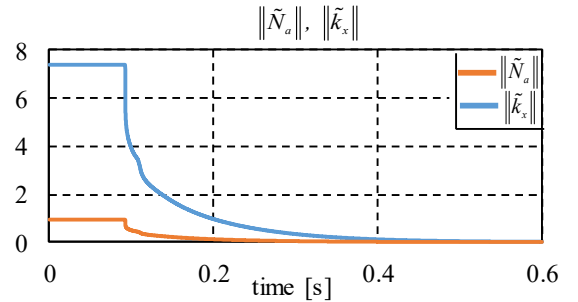

Figure 1. Transient performance of the norm of errors $\tilde{N}_a$ and $\tilde{k}_x$.

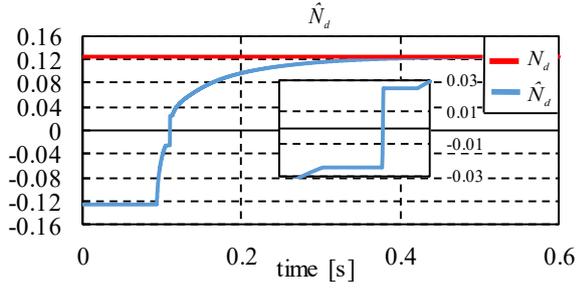

Figure 2. Comparison of values of ideal $N_d$ and its estimation.

Having analyzed Fig.1 and Fig.2 with the help of (30), it is concluded that the exponential convergence to the ideal values was provided for $\hat{k}_x$ and $\hat{k}_r$. One switching (31) happened in the course of transients. This validated the theoretical results shown in Proposition 2 and Theorem 1. Figure 3 shows the comparison of the transients of the state vectors elements of both the plant and the reference model. It was concluded that the tracking error $e_{ref}$ converged to zero. This validated the results presented as Theorem 1.

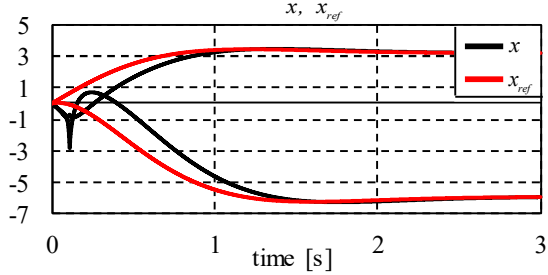

Figure 3. Transients of state vectors elements of plant and reference model.

## IV. CONCLUSION

Developed I-DREM MRAC included the control law (3), the law of its parameters adjustment (32), analytical equation (30) and the non-linear operator (31). Such method can be applied to control a generic class of multi-input, multi-output (MIMO) LTI systems, which $B$ matrix is of full column rank (aircraft control systems, robotic arms and different process control systems). In contrast to the conventional MRAC, the I-DREM MRAC method: 1) provided exponential convergence of the adjustable parameters when $\omega \in$ IE; 2) included the adjustment law of the adaptation rate matrix $\Gamma$; 3) did not require knowledge of the plant $B$ matrix. However, since the filter output (22) becomes insensitive to the new input data after $(3...5) \cdot \sigma^{-1}$ seconds, the application of the I-DREM MRAC is only possible if the following assumption is satisfied.

***Assumption 2.*** *The ideal values of the control law parameters $\forall t \geq 0$ are such that $k_x = const$ and $k_r = const$.*

This assumption, in contrast to the MRAC standard assumption on the quasi-stationarity of the controller parameters, do not allow $k_x$ and/or $k_r$ to be quasi-stationary or defined as intervals. Therefore, the scope of future research is to present the results of the filter (22) modification to relax Assumption 2 to the conventional MRAC one of the controller parameters quasi-stationarity.

## APPENDIX

### A. Proof of Lemma 1

The solution of the differential equation (10) is written as:

$$\mu_f(t) = e^{-l \cdot t}\mu_f(0) + e^{-l \cdot t}\int_0^t e^{l \cdot \tau}\dot{e}_{ref}(\tau)d\tau; \quad \text{(A1)}$$

$$e_f(t) = e^{-l \cdot t}e_f(0) + e^{-l \cdot t}\int_0^t e^{l \cdot \tau}e_{ref}(\tau)d\tau.$$

Using the method of integration by parts, the following integral from the $\mu_f(t)$ equation is taken:

$$\int_0^t e^{l \cdot \tau}\dot{e}_{ref}(\tau)d\tau \equiv \begin{bmatrix} u = e^{l \cdot \tau};\ du = le^{l \cdot \tau}d\tau \\ dv = \dot{e}_{ref}(\tau)\ d\tau;\ v = e_{ref}(\tau) \end{bmatrix} =$$
$$= uv\big|_0^t - \int_0^t v\ du = e^{l \cdot \tau}e_{ref}(\tau)\big|_0^t - l\int_0^t e_{ref}(\tau)e^{l \cdot \tau}d\tau = \quad \text{(A2)}$$
$$= e^{l \cdot t}e_{ref}(t) - e_{ref}(0) - l\int_0^t e_{ref}(\tau)e^{l \cdot \tau}d\tau.$$

Considering the equation for $e_f(t)$, the obtained solution is substituted into the equation for $\mu_f(t)$:

$$\mu_f(t) = e^{-l \cdot t}\mu_f(0) + e_{ref}(t) - e^{-l \cdot t}e_{ref}(0) - $$
$$- e^{-l \cdot t}l\int_0^t e_{ref}(\tau)e^{l \cdot \tau}d\tau = e^{-l \cdot t}\mu_f(0) + e_{ref}(t) - \quad \text{(A3)}$$
$$- e^{-l \cdot t}e_{ref}(0) - le_f(t) + le^{-l \cdot t}e_f(0).$$

It follows from (A3) that $\mu_f$ can be calculated as (12). This completes the proof.

### B. Proof of Theorem 1

The convergence of the constituent parts of the vector ξ will be shown to prove that the whole vector also converges to zero. The Lyapunov function candidate to study the convergence of the parameter error $\tilde{\theta}_{vec}$ is chosen as:

$$V = \tilde{\theta}_{vec}^T \Gamma^{-1} \tilde{\theta}_{vec}. \quad \text{(B4)}$$

Let the results of [20] be used, where the boundedness of the adaptation rate $\Gamma$ was proved for the regressor $\Omega(t)$, which had the properties described in Proposition 1. Then the function (B4) is bounded:

$$\lambda_{min}(\Gamma^{-1})\|\tilde{\theta}_{vec}\|^2 \leq \|V\| \leq \lambda_{max}(\Gamma^{-1})\|\tilde{\theta}_{vec}\|^2. \quad \text{(B5)}$$

The derivative of (B4) is written as:

$$\dot{V} = 2\tilde{\theta}_{vec}^T \Gamma^{-1}\dot{\tilde{\theta}}_{vec} + \tilde{\theta}_{vec}^T \dot{\Gamma}^{-1}\tilde{\theta}_{vec}. \quad \text{(B6)}$$

The following equation is used to obtain the adaptation law for $\Gamma^{-1}$ from the one for $\Gamma$:

$$\frac{d}{dt}I = \frac{d}{dt}\left[\Gamma^{-1}\Gamma\right] = \Gamma\frac{d\Gamma^{-1}}{dt} + \Gamma^{-1}\frac{d\Gamma}{dt} = 0. \quad \text{(B7)}$$

Considering (B7) and (32), (B6) is rewritten as:

$$\dot{V} = -2\tilde{\theta}_{vec}^T \Omega^{2m} \tilde{\theta}_{vec} + \tilde{\theta}_{vec}^T \left[ \Omega^{2m} - \lambda \Gamma^{-1} \right] \tilde{\theta}_{vec} = \\ = -\tilde{\theta}_{vec}^T \Omega^{2m} \tilde{\theta}_{vec} - \tilde{\theta}_{vec}^T \lambda \Gamma^{-1} \tilde{\theta}_{vec}. \quad \text{(B8)}$$

The lower bound of the regressor $\Omega(t)$ will be considered to obtain the lower bound of (B8). Considering Proposition 1, for $\Omega(t)$ $\exists T > t_0$ such that:

$$\Omega(t) = \int_{t_0}^{t} e^{-\sigma\tau} \omega^2(\tau) d\tau \geq \int_{t_0}^{T} e^{-\sigma\tau} \omega^2(\tau) d\tau = \Omega(T). \quad \text{(B9)}$$

Then $\forall t \geq T$ $\Omega^{2m}(t) \geq \Omega^{2m}(T)$. So the lower bound of (B8) is written as:

$$\dot{V} \leq -\left[ \Omega^{2m}(T) + \lambda \lambda_{min}(\Gamma^{-1}) \right] \|\tilde{\theta}_{vec}\|^2 = -\kappa V, \\ \kappa = \left[ \Omega^{2m}(T) + \lambda \lambda_{min}(\Gamma^{-1}) \right] \lambda_{max}^{-1}(\Gamma^{-1}). \quad \text{(B10)}$$

The differential inequality (B10) is solved $\forall t \geq T$:

$$\|\tilde{\theta}_{vec}\| \leq \sqrt{\lambda_{min}^{-1}(\Gamma^{-1}) e^{-\kappa(t-T)} V(T)}. \quad \text{(B11)}$$

Then $\tilde{\theta}_{vec}$ converges to zero exponentially $\forall t \geq T$. So $\forall t \geq (3\ldots5)\kappa^{-1}$ the error equation (9) can be written as:

$$\dot{e}_{ref} = A_{ref} e_{ref}; \quad e_{ref}(0) = e_{ref}\left([3\ldots5]\kappa^{-1}\right). \quad \text{(B12)}$$

The Lyapunov function candidate to study the convergence of the tracking error (B12) $\forall t \geq (3\ldots5)\kappa^{-1}$ is chosen as:

$$L = e_{ref}^T e_{ref}. \quad \text{(B13)}$$

The derivative of (B13) is written as (B14) as $A_{ref}$ is the Hurwitz matrix.

$$\dot{L} = 2 e_{ref}^T A_{ref} e_{ref} \leq -2\lambda_{min}(A_{ref}) \|e_{ref}\|^2 = -\kappa_{e_{ref}} L \quad \text{(B14)}$$

The inequality (B14) is solved $\forall t \geq (3\ldots5)\kappa^{-1}$:

$$\|e_{ref}\| \leq \sqrt{e^{-\kappa_{e_{ref}}(t-[3\ldots5]\kappa^{-1})} L\left([3\ldots5]\kappa^{-1}\right)}. \quad \text{(B15)}$$

Then $e_{ref}$ converges exponentially $\forall t \geq (3\ldots5)\kappa^{-1}$. Combining the inequalities (B11) and (B15), the following inequality for the whole vector $\xi$ is obtained:

$$\|\xi\| \leq \sqrt{e^{-\eta(t-T_\xi)} V_\xi}, \\ \eta = \min\{\kappa, \kappa_{e_{ref}}\}; \quad T_\xi = \max\{T, [3\ldots5]\kappa^{-1}\}, \\ V_\xi = \max\{\lambda_{min}^{-1}(\Gamma^{-1}) V(T), L([3\ldots5]\kappa^{-1})\}. \quad \text{(B16)}$$

So the error $\xi$ exponentially converges to zero with the rate, which can be made arbitrarily high by choice of the forgetting factor $\lambda$ of the $\Gamma$ adaptation law or the memory factor $\sigma$ of the filter (22). This completes the proof of Theorem 1.

## REFERENCES

[1] K. S. Narendra, and A. M. Annaswamy, *Stable adaptive systems*. Courier Corporation, 2012.

[2] P. A. Ioannou, J. Sun, *Robust adaptive control*. Courier Corporation, 2012.

[3] B. Anderson, "Exponential stability of linear equations arising in adaptive identification," *IEEE Transactions on Automatic Control*, vol. 22, no. 1, pp. 83-88, 1977.

[4] J. S. Lin, and I. Kanellakopoulos, "Nonlinearities enhance parameter convergence in output-feedback systems," *IEEE Transactions on Automatic Control*, vol. 43, no. 2, pp 204-222, 1998.

[5] B. M. Jenkins, A. M. Annaswamy, E. Lavretsky, and T. E. Gibson, "Convergence properties of adaptive systems and the definition of exponential stability," *SIAM journal on control and optimization*, vol. 56, no. 4, pp. 2463-2484, 2018.

[6] P. A. Ioannou, and P. V. Kokotovic, "Instability analysis and improvement of robustness of adaptive control," *Automatica*, vol. 20, no. 5, pp. 583-594, 1984.

[7] S. Boyd, and S. S. Sastry, "Necessary and sufficient conditions for parameter convergence in adaptive control," *Automatica*, vol. 22, no. 6, pp. 629-639, 1986.

[8] S. B. Roy, and S. Bhasin, "Switched MRAC with improved performance using semi-initial excitation," in *Proc. 2018 Annual American Control Conference*, 2018, pp. 74-79.

[9] S. B. Roy, S. Bhasin, and I. N. Kar, "Combined MRAC for unknown MIMO LTI systems with parameter convergence," *IEEE Transactions on Automatic Control*, vol. 63, no. 1, pp. 283-290, 2017.

[10] G. Chowdhary, T. Yucelen, M. Muhlegg, and E. Johnson, "Concurrent learning adaptive control of linear systems with exponentially convergent bounds," *International Journal of Adaptive Control and Signal Processing*, vol. 27, no. 4, pp. 280-301, 2013.

[11] R. Ortega, V. Nikiforov, and D. Gerasimov, "On modified parameter estimators for identification and adaptive control: a unified framework and some new schemes," *Annual Reviews in Control*, to be published. DOI: 10.1016/j.arcontrol.2020.06.002.

[12] D. N. Gerasimov, R. Ortega, and V. O. Nikiforov, "Relaxing the high-frequency gain sign assumption in direct model reference adaptive control," *European Journal of Control*, vol. 43, pp. 12-19, 2018.

[13] S. Aranovskiy, A. Bobtsov, R. Ortega, and A. Pyrkin, "Performance enhancement of parameter estimators via dynamic regressor extension and mixing," *IEEE Transactions on Automatic Control*, vol. 62, no. 7, pp. 3546-3550, 2016.

[14] K. A. Wise, E. Lavretsky, and N. Hovakimyan, "Adaptive control of flight: theory, applications, and open problems," in *Proc. 2006 American Control Conference*, 2006, pp. 1-6.

[15] T. Yucelen, and W. M. Haddad, "Low-frequency learning and fast adaptation in model reference adaptive control," *IEEE Transactions on Automatic Control*, vol. 58. no. 4. pp. 1080-1085, 2012.

[16] N. Hovakimyan, and C. Cao, *$L_1$ adaptive control theory: Guaranteed robustness with fast adaptation*. Philadelphia, PA: Society for Industrial and Applied Mathematics, 2010.

[17] S. P. Schatz, T. Yucelen, B. Gruenwal, and F. Holzapfe, "Application of a novel scalability notion in adaptive control to various adaptive control frameworks," in *Proc. AIAA Guidance, Navigation, and Control Conference*, 2015, pp. 1-17.

[18] J. Jaramillo, T. Yucelen, and K. Wilcher, "Scalability in Model Reference Adaptive Control," *AIAA Scitech 2020*, 2020, pp. 1-13.

[19] N. T. Nguyen, *Model-Reference Adaptive Control: A Primer*. Springer, 2018.

[20] A. Glushchenko, V. Petrov, and K. Lastochkin, "Robust method to provide exponential convergence of model parameters solving LTI plant identification problem," *arXiv preprint arXiv:2009.14496*, pp. 1-18, 2020.

[21] K. S. Narendra, and P. Kudva, "Stable adaptive schemes for system identification and control – Part I," *IEEE Transactions on Systems, Man, and Cybernetics*, vol.6., pp. 542-551, 1974.

[22] B. Yi, and R. Ortega, "Conditions for Convergence of Dynamic Regressor Extension and Mixing Parameter Estimator Using LTI Filters," *arXiv preprint arXiv:2007.15224*, pp. 1-6, 2020.